\documentclass[conference,compsoc]{IEEEtran}
\IEEEoverridecommandlockouts
\usepackage{cite}
\usepackage{amsmath,amssymb,amsfonts}
\usepackage{algorithmic}
\usepackage{graphicx}
\usepackage{textcomp}
\usepackage{xcolor}
\def\BibTeX{{\rm B\kern-.05em{\sc i\kern-.025em b}\kern-.08em
    T\kern-.1667em\lower.7ex\hbox{E}\kern-.125emX}}

\usepackage{array}
\usepackage{multirow}
\usepackage{threeparttable}

\usepackage{makecell}
\usepackage{hyperref}
\usepackage[square,numbers]{natbib}
\usepackage{booktabs}
\graphicspath{{figure/}}
\usepackage{xcolor}
\usepackage{verbatim}
\hypersetup{hidelinks}

\begin{document}


\title{Pedagogical Friction in Password Creation}

\author{%
\centering
\renewcommand{\arraystretch}{1.8}
\begin{tabular}{ccc}

\begin{minipage}[t]{0.31\textwidth}
\centering
\textbf{Qian Ma}\\
\textit{The Pennsylvania State University}\\
qfm5033@psu.edu
\end{minipage}
&
\begin{minipage}[t]{0.31\textwidth}
\centering
\textbf{Yingfan Zhou}\\
\textit{The Pennsylvania State University}\\
yxz5975@psu.edu
\end{minipage}
&
\begin{minipage}[t]{0.31\textwidth}
\centering
\textbf{Shubhang Kaushik}\\
\textit{The Pennsylvania State University}\\
smp7224@psu.edu
\end{minipage}
\\[2em]

\begin{minipage}[t]{0.31\textwidth}
\centering
\textbf{Aamod Joshi}\\
\textit{The Pennsylvania State University}\\
avj5440@psu.edu
\end{minipage}
&
\begin{minipage}[t]{0.31\textwidth}
\centering
\textbf{Aditya Majumdar}\\
\textit{The Pennsylvania State University}\\
adity@psu.edu
\end{minipage}
&
\begin{minipage}[t]{0.31\textwidth}
\centering
\textbf{Noah Apthorpe}\\
\textit{Colgate University}\\
napthorpe@colgate.edu
\end{minipage}
\\[2pt]

\begin{minipage}[t]{0.31\textwidth}
\centering
\textbf{Yan Shvartzshnaider}\\
\textit{York University}\\
yansh@yorku.ca
\end{minipage}
&
\begin{minipage}[t]{0.31\textwidth}
\centering
\textbf{Sarah Rajtmajer}\\
\textit{The Pennsylvania State University}\\
smr48@psu.edu
\end{minipage}
&
\begin{minipage}[t]{0.31\textwidth}
\centering
\textbf{Brett Frischmann}\\
\textit{Villanova University}\\
brett.frischmann@law.villanova.edu
\end{minipage}

\end{tabular}
}

\maketitle

\begin{abstract}
Users often make security- and privacy-relevant decisions without a clear understanding of the rules that govern safe behavior. We introduce pedagogical friction, a design approach that inserts brief, instructional interactions at the moment of action. We evaluate this approach in the context of password creation, a familiar task with clear quality criteria. 

We conducted a randomized study with 128 participants across four conditions varying the depth and interactivity of guidance. We assessed three outcomes: (1) rule compliance in a subsequent password task without guidance; (2) accuracy on quiz questions tied to password rules; and (3) behavior-knowledge alignment, which captures whether participants who correctly followed a rule also recognized it on the quiz. Across guided conditions, participants corrected most rule violations in the follow-up task and showed high behavior-knowledge alignment. Quiz accuracy was significantly higher for the interactive condition than the control on symbol-related questions after correction. Our results position pedagogical friction as a lightweight intervention for security- and privacy-critical interfaces.
\end{abstract}


\section{Introduction}
\noindent Users frequently encounter privacy- and security-critical decisions without clear guidance about how to make safe choices. Interfaces commonly present these actions as simple form fields, offering only minimal cues about what constitutes a good input or why it matters. 

We propose \textbf{pedagogical friction}---a design framework for inserting brief instructional moments at the point of action. The framework is informed by prior work on persuasive technology, microlearning, and friction in design. Rather than separating guidance from behavior through training or retrospective warnings, pedagogical friction delivers explanations when users can immediately act on them~\cite{DBLP:journals/ubiquity/Fogg02,DBLP:conf/persuasive/Fogg09a}. We study this in the context of password creation.

Passwords offer an ideal testbed for several reasons. First, password strength admits objective measurement through adversarial guessing models~\cite{DBLP:conf/uss/MelicherUSKBCC16,DBLP:conf/ccs/MazurekKVBCCKSU13}, providing clear ground truth for evaluating interventions. Second, passwords are ubiquitous. The task requires no specialized domain knowledge, enabling engagement with general populations. Yet there is a sizable gap between lay and expert knowledge about password security~\cite{frischmann2023common}. Third, despite decades of research attention and policy guidance, people continue to create predictable passwords, recycle credentials across accounts, and fall back on patterns tied to personal information~\cite{DBLP:conf/soups/UrNBSSBCC15,moh2024understanding,DBLP:conf/soups/ZhangPBC19}. These behaviors leave accounts vulnerable to credential stuffing, brute-force attacks, and phishing~\cite{wang2020detecting,simon2025response}.

Because many accounts contain sensitive personal information~\cite{mayer2023awareness,kocabas2021understanding}, weak or reused passwords have direct privacy consequences. Compromise can enable undetected account access and modification~\cite{sahin2025you}. Even when services use strong protections, those protections still rely on passwords that are hard to guess or reuse~\cite{blessing2025sok,wei2025trying}. In this sense, we view pedagogical friction as a human-centered privacy intervention at the moment users create their passwords.

Widely used interventions to improve password strength, such as composition rules, periodic resets, and real time strength meters often lead to frustration, mechanical compliance, and little lasting benefit~\cite{DBLP:conf/soups/LeeSN22,DBLP:journals/ieeesp/YanBAG04}. Users learn to satisfy validators with superficial edits rather than adopt genuinely stronger strategies, a pattern observed in studies of strength meter use and feedback tuning~\cite{DBLP:conf/ccs/MazurekKVBCCKSU13,DBLP:journals/tissec/StobertB18,DBLP:conf/ccs/ChiassonFSOB09}. These outcomes reflect a long standing tension between usability and security, where stricter controls can increase effort and encourage workarounds~\cite{hielscher2023lacking,DBLP:conf/chi/EgelmanCH08}. 

Pedagogical friction seeks to address this by operationalizing principles from microlearning~\cite{hug2006microlearning,allela2021introduction,sankaranarayanan2023microlearning}, where short, focused guidance is embedded within a task. It aligns with value sensitive design ~\cite{friedman2013value}, which treats user goals and constraints as first class considerations in interface decisions. It also draws on behavioral psychology, where timely cues increase attention and make corrective action more likely~\cite{DBLP:journals/ubiquity/Fogg02}. We ask: \emph{Can pedagogical friction support in situ learning of best practices for password generation?}

We investigate this question using a randomized experiment with 128 participants across four interface conditions varying the depth and interactivity of guidance. Specifically, during password creation, participants are presented with: (T0) \emph{meter-only control} (red/yellow/green reflecting password strength); (T1) \emph{brief tips}, conveying simple rules for creating strong passwords; (T2) \emph{detailed tips}, giving rules with explanations; or, (T3) \emph{interactive tips} (users required to click to acknowledge). We log each password attempt via keystrokes, edits, and timing. This is followed by a 5-minute survey of demographics, digital literacy, and privacy attitudes. Following the survey, participants are engaged in  another meter-only password task followed by a brief quiz to test their knowledge of best practices for password creation. 

We study learning through three outcomes: tip compliance in the post-test phase; accuracy on quiz questions tied to password tips, and behavior-knowledge alignment, comparing adherence to best practices (following rules for creating strong passwords) and explicit recognition of these rules. Across guided conditions, participants corrected most rule violations flagged during guidance and showed high behavior-knowledge alignment. Quiz results showed the clearest advantage for symbol-related questions, where the interactive condition significantly outperformed the control after correction. In practice, this suggests that designers may not need heavy handed interventions to influence secure behavior in password creation, although learning gains may be stronger for some rule types than for others. Our work contributes:
\begin{enumerate}

\item \textbf{a design framework}, \emph{pedagogical friction}, which we define as brief, rule-linked prompts that add minimal effort while supporting learning during password creation;

\item \textbf{results from an experiment} varying the depth and interactivity of guidance during password creation and examining effects on rule compliance, rule-linked quiz accuracy, and behavior-knowledge alignment; and
\item \textbf{a rule-centered measurement framework} with group-, user-, and rule-level compliance and alignment measures.

\end{enumerate}

\section{Related Work}

\subsection{Friction in Design}

\noindent Work in human-computer interaction (HCI), communication, and law has examined different forms of prosocial friction in interfaces and infrastructures~\cite{cox2016design,natali2023per,frischmann2023friction,cabitza2019programmed,gruning2024framework,chalmers2003seamful,grosse2013slow}. Terminology has varied across disciplines, including design frictions, frictional design, friction-in-design, microboundaries, programmed inefficiencies, seamful design, and slow design. In HCI, seamful design and related work deliberately reveal breakdowns in the underlying infrastructure as resources for awareness and reflection~\cite{chalmers2003seamful}. \citet{cox2016design} argue that thoughtfully-introduced microboundaries and design frictions can interrupt mindless, automatic interactions, prompting moments of reflection and more mindful decision-making about ongoing use.

\begin{figure}[t]
  \centering
  \includegraphics[width=\linewidth]{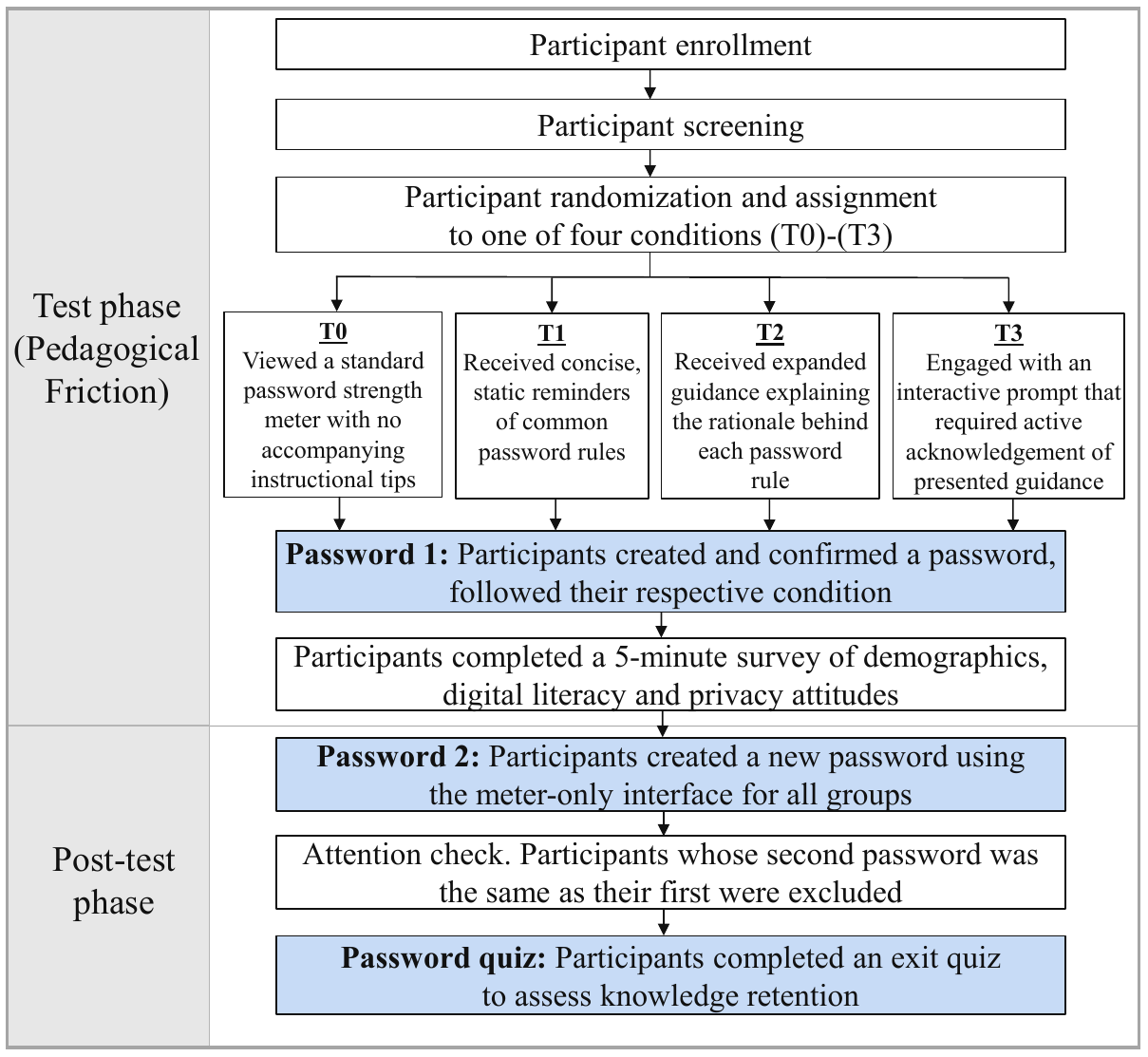}
  \caption{Study procedure. In the test phase, participants generate a password within one of four conditions and complete a survey of demographics, digital literacy and privacy attitudes. In the post-test phase, participants complete a second meter-only password creation task and take a quiz on best practices for password creation.}
  \label{fig:flowchart}
\end{figure}

Frischmann and Benesch~\cite{frischmann2023friction} argue that some resistance which slows people down can support reflection, experimentation, and self-determination. They propose a framework for analyzing friction-in-design along parameters such as type of friction, direct effects on subjects, and intended purposes and impacts. They highlight security and safety examples in which more active warnings that interrupt users and require interaction can improve outcomes compared to passive, easily ignored indicators. This pattern is echoed in empirical work on security warnings, where interstitial warnings that interrupt the task and require user interaction are substantially more effective at changing behavior than contextual warnings that remain in the background~\cite{kaiser2021adapting}. Likewise, related work in security and privacy finds that prominent notifications and other active channels can change user actions more effectively than less visible cues~\cite{utz2023comparing}.

Our intervention follows these ideas by using brief, embedded prompts that add a small amount of friction to the password entry workflow. During the test phase, participants in the guided conditions receive rule specific guidance, with T3 additionally requiring acknowledgment before proceeding. We then measure effects on rule compliance, recall, and understanding in the post-test phase.

\subsection{Microlearning}

\noindent Microlearning is a relatively broad term referring to instructional approaches that deliver targeted, action-oriented, bite-sized content to achieve specific objectives within a short period, typically within a few seconds or minutes \cite{monib2025microlearning,cronin2024microlearning}. Prior work suggests that microlearning supports better knowledge and skill acquisition compared with some traditional methods \cite{de2019microlearning, cronin2024microlearning}. Instruction that appears inside a task can focus attention, prompt immediate correction, and support retention more effectively than detached training~\cite{DBLP:conf/chi/EgelmanP15}. This approach aligns with behavioral models showing that secure actions occur when motivation, ability, and a timely prompt converge~\cite{DBLP:conf/persuasive/Fogg09a}. 

In our design, the password interface itself delivers that prompt. It provides feedback at the moment when users are able and motivated to revise their input. 

\begin{figure*}[t]
  \centering
  \includegraphics[width=\linewidth]{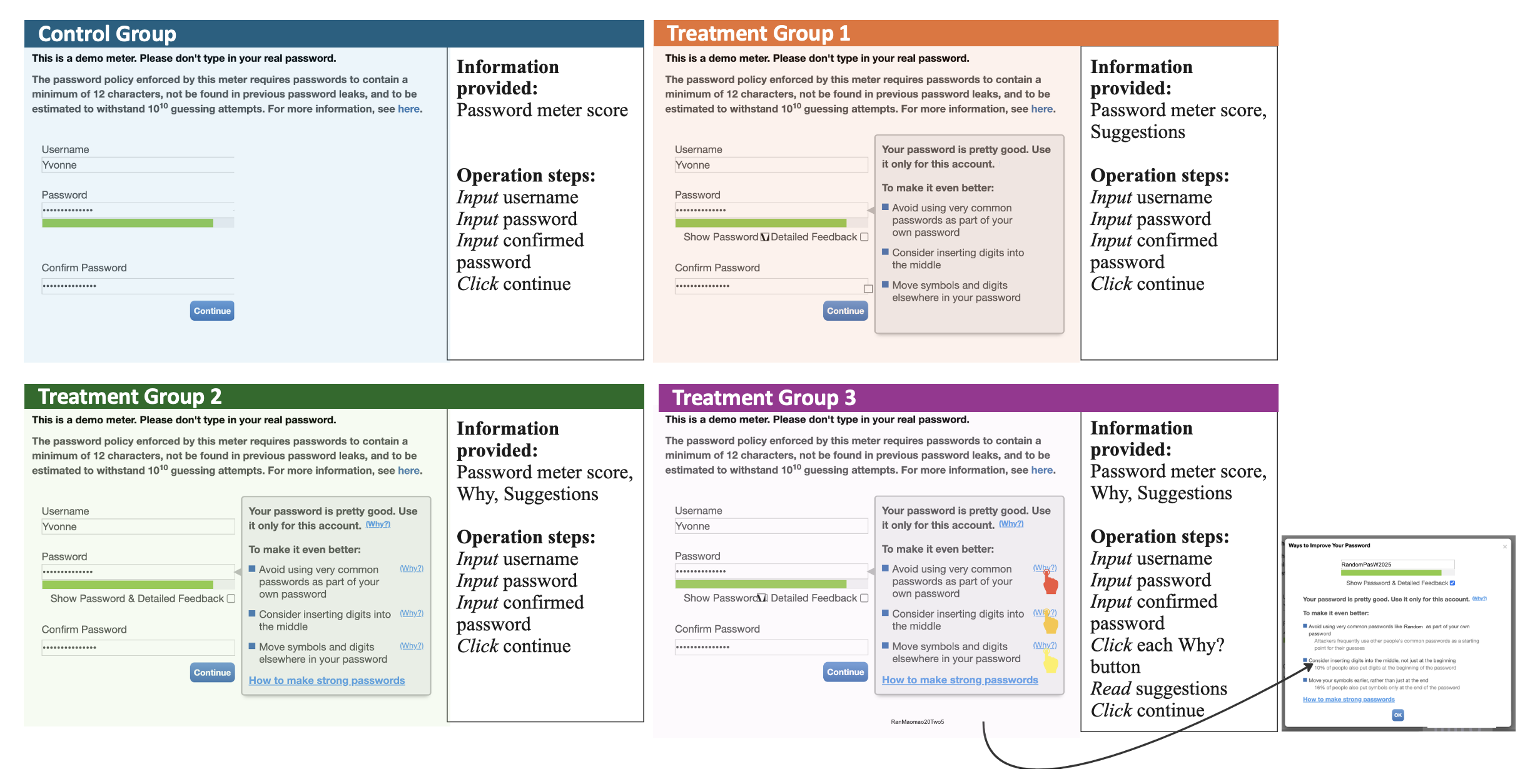}
  \caption{Four password creation interfaces varying in instructional guidance: (T0) meter-only control; (T1) brief tips; (T2) detailed tips; and (T3) interactive tips requiring acknowledgment.} 
  \label{Fig:expt}
\end{figure*}

\subsection{Password Security}

\noindent Although researchers and industry groups increasingly advocate for passwordless authentication, passwords remain deeply embedded in everyday sign-in workflows and are still the most common form of digital authentication~\cite{lassak2024aren,merdenyan2022two,DBLP:conf/ccs/GollaD18}. Their persistence reflects practical advantages such as conceptual simplicity, cost efficiency, and cross platform compatibility~\cite{farke2020you,DBLP:conf/sp/BonneauHOS12}. Consequently, improving how users learn to construct stronger passwords remains a central challenge in usable security research~\cite{amador2023prospects,wash2016understanding}.

Despite repeated warnings and policy updates, risky password habits remain common across user populations with different technical backgrounds~\cite{moh2024understanding,DBLP:conf/soups/ZhangPBC19}. Studies of adaptive and personalized security interventions show that systems which adjust the intensity or form of feedback outperform one-size-fits-all designs~\cite{schoni2025s,zheng2023checking,ur2012smart}. In parallel, brief task-embedded interruptions can influence user decisions: interstitial warnings that require active acknowledgment have been found to outperform passive contextual notices~\cite{kaiser2021adapting}. Gamified learning tools can also raise motivation and engagement in cybersecurity education when feedback is immediate and progress is visible~\cite{gwenhure2025gamified,van2021successful}. Longitudinal work further shows that repeated exposure and reinforcement are important for retaining secure behaviors over time~\cite{toth2025sustaining,DBLP:conf/huc/LazarKTN15}. Guided by these findings, our study uses graded levels of friction during password creation and includes a follow-up test of recall and understanding.

\section{Method}
\label{sec:method}

\noindent We conducted a randomized experiment in which password creation tasks were embedded within a multi-step protocol (see Fig.~\ref{fig:flowchart} for an overview). 

To preserve study validity, we used mild deception. Participants were told they were enrolling in a survey study of digital literacy and privacy attitudes, and that a password was required to ``log in'' to the survey platform. This framing concealed the fact that password creation tasks were the focus of the experiment. No false information was provided about risks, data handling, or compensation, and the deception did not alter participants’ rights or welfare. The protocol, including the use of deception, received prior approval from our Institutional Review Board (IRB) and participants were debriefed following completion of the study.

\subsection{Experimental Conditions}

\noindent Following enrollment and screening, participants were randomly assigned to one of four experimental conditions (T0, T1, T2, T3), which differed in the depth and form of instructional guidance during password creation. Screenshots of the four interfaces are provided in Fig.~\ref{Fig:expt}. The interfaces were created using the method and publicly available code from Ur et al.\ and the associated Password Guessability Service~\cite{DBLP:conf/chi/UrAABCCCDNHJM17,pgs-service}.

\noindent \textbf{Control Group (T0) - Meter only}.
Participants viewed a standard password strength meter (green/yellow/red bar) with no accompanying instructional tips. 

The password meter in our study is treated as evaluative feedback rather than pedagogical friction. It indicates password strength, but does not provide the rule-linked instruction or active acknowledgment. 

\noindent \textbf{Treatment Group 1 (T1) - Brief tips}.
Participants received concise feedback related to password best practices (\emph{rules}) (e.g., ``Add symbols in unpredictable locations'' and ``Avoid using a pattern on your keyboard'') displayed passively below the password field. This feedback (\emph{tips}) were triggered dynamically based on a participant's input string. 

As described in \cite{DBLP:conf/chi/UrAABCCCDNHJM17}, users' input could trigger one or more of 20 possible tips reflecting best practices for password selection (see Appendix Table~\ref{tab:survey_questions} for full listing).

\noindent \textbf{Treatment Group 2 (T2) - Detailed tips}.
As in T1, tips were triggered dynamically based on a participant's input. However, in this condition, participants were shown expanded tips that explained the rationale behind each password rule, offering brief context on why certain practices improve strength. These tips appeared prominently but required no user interaction.

\noindent \textbf{Treatment Group 3 (T3) - Interactive tips}.
Building upon T2, participants in T3 were required to engage with an interactive prompt that required acknowledgment of understanding of at least one tip and its rationale before proceeding. Specifically, participants were required to click on at least one ``Why?'' prompt before proceeding (see lower right quadrant of Fig.~\ref{Fig:expt}). This condition incorporated the highest level of friction by requiring active reflection on the guidance presented.

This tiered design allows systematic comparison across increasing pedagogical friction.

\subsection{Procedure}

\noindent \textbf{Test phase (pedagogical friction).} \textbf{(1)} Participants were placed into their assigned condition and instructed accordingly to create and confirm a password in order to access the survey. 
Participant interactions with the password interface were recorded in real time, including typing duration, number of edits, and tip interactions (for T3).  

\noindent   \textbf{(2)} Participants then completed an approximately 5-minute survey of demographics, digital literacy, and privacy attitudes, which served to provide contextual data for our analyses as well as a cover story (mild deception) for password tasks.\\

\noindent \textbf{Post-test phase.} \textbf{(3)} After completing (2), participants were asked to create a new password. This second password task used the meter-only interface and was presented in the same way for all groups (T0)-(T3).

We aim to assess whether prior exposure to tips or friction influenced subsequent password behavior. Participants were clearly instructed to create a \emph{different} password than the one they created earlier for login.

\noindent   \textbf{(4)} Finally, all participants then completed the same knowledge retention quiz, which included multiple-choice questions, some of which were directly tied to the instructional tips shown in the test phase.

\subsection{Participants and Exclusions}
\noindent Participants were recruited via Prolific.\footnote{\url{https://www.prolific.com}} Eligibility required that participants be at least 18 years old. Sessions lasted approximately 10 minutes, and participants were paid \$15 per hour, pro-rated to their completion time.

Reusing the same password across both phases was treated as a failure to follow explicit study instructions, and those users were excluded as a post-test attention check. The final analytic dataset comprised $128$ participants, distributed across the four groups as follows: T0 ($47$), T1 ($22$), T2 ($29$), and T3 ($30$). 

We conducted a priori power analysis in G\textasteriskcentered Power assuming a medium size between-group effect ($f = 0.3$), $\alpha = 0.05$, and power $= 0.8$. This analysis indicated that an analytic sample of 128 participants would provide reasonable power for the planned group comparisons; we treat smaller rule-specific subgroup analyses as exploratory.

\begin{table}[t]
\centering
\caption{Password rules introduced through tips}
\label{tab:rule_set}
\fontsize{7}{8.5}\selectfont
\begin{tabular}{
    >{\centering\arraybackslash}p{1.5cm} |
    >{\raggedright\arraybackslash}p{5.5cm}}
\toprule
\textbf{Rule} & \textbf{Definition as presented in the study} \\
\midrule
\textit{Length} & Password should meet or exceed the minimum length required ($\geq$\,8 characters). \\
\midrule
\textit{Symbol} & Include at least one non-alphanumeric character. \\
\midrule
\textit{Uppercase} & Include at least one uppercase letter. \\
\midrule
\textit{Number} & Include at least one digit. \\
\midrule
\textit{\makecell{Personal Info}} & Avoid personal identifiers (e.g., names, usernames, email handles). \\
\midrule
\textit{Dictionary} & Avoid common dictionary words. \\
\bottomrule
\end{tabular}
\end{table}

\subsection{Demographics}

Participant demographics are summarized in Appendix Table~\ref{tab:demographics}. The sample was 53.1\% female, 43.8\% male. Participants ranged in age from 18 to 65 years and older. The sample was predominantly White (61.7\%), followed by Black (21.9\%). Educational level was relatively high, with 42.2\% holding a bachelor's degree. 

We used chi-square tests of homogeneity to examine demographic balance across the four groups. No significant differences were detected across groups for any demographic variable (all \emph{p}~$\geq$~0.05), indicating that random assignment produced comparable participant samples.

\begin{figure*}[t]
    \centering
    \includegraphics[width=0.82\linewidth,height=4cm]{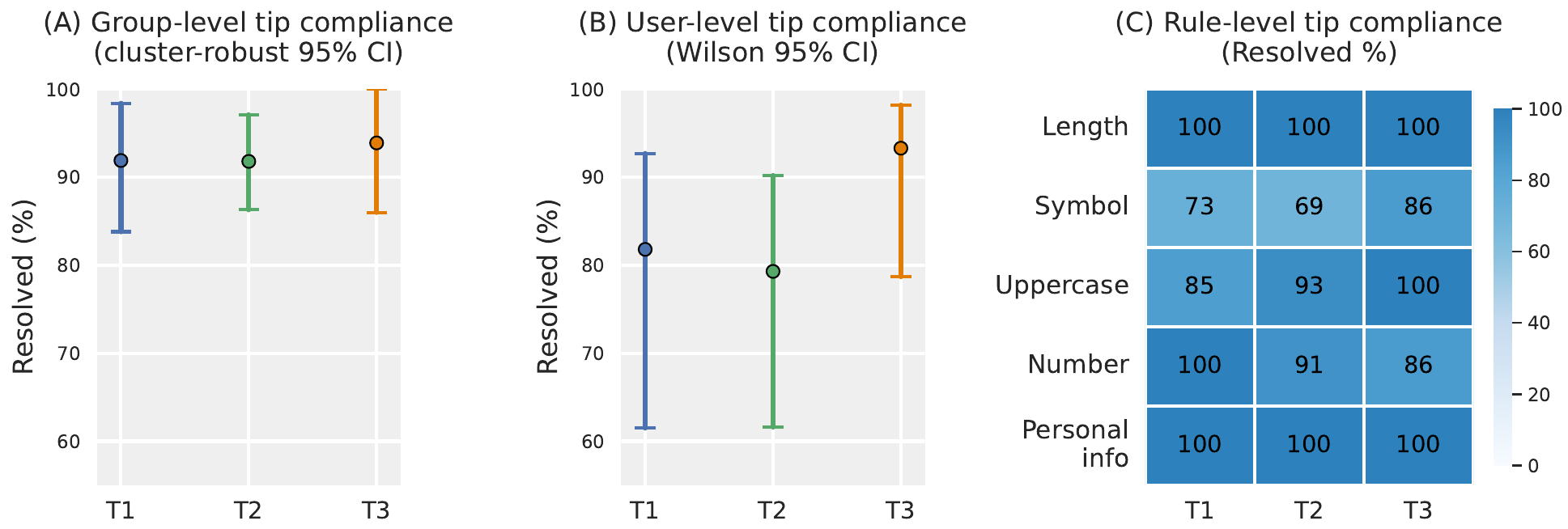}
    \caption{(A) Group-level tip compliance. (B) User-level tip compliance. (C) Rule-level tip compliance. (all rules resolved)}
    \label{fig:tip_compliance}
\end{figure*}

\subsection{Instrumentation and Measures}
\label{subsec:instrumentation_measures}
\noindent Password creation events were recorded through a custom web platform capturing keystrokes, timestamps and metadata. These records allow us to derive interaction telemetry as well as rule-linked behavioral and knowledge measures.

\subsubsection{Interaction telemetry} Telemetry variables help to contextualize differences in engagement across conditions but are not treated as primary outcome measures. 

\begin{itemize}
\item \textbf{Edit count}. Total character editing actions during password entry, including insertions and deletions.

\item \textbf{Tip interactions (T3 only)}. Counts of acknowledgment actions (clicking on ``Why?'') within the interactive tip interface. This measure reflects the extent of engagement with the highest friction condition. 
\end{itemize}

\subsubsection{Rule categories} 
For rule-based analyses, we mapped 20 heuristic rules and tips, password violations, and quiz questions to six rule categories (\textit{Length}, \textit{Symbol}, \textit{Uppercase}, \textit{Number}, \textit{Personal information}, \textit{Dictionary}; see Table~\ref{tab:rule_set} for the six rules and Appendix Table~\ref{tab:rule_mappping} for the mapping from quiz target rules to six rule categories). For participants in guided conditions, rules that were violated in the test phase and triggered at least one tip formed that participant’s post-test rule set. Each quiz question was linked to a unique rule (see Appendix Table~\ref{tab:survey_questions} for the full quiz instrument). 

\subsubsection{Compliance measures} 
Our analyses focus on two complementary constructs: 

\begin{itemize}
\item \textbf{Tip compliance}. Whether participants complied with the tips they were shown (\emph{corrected} rule violations) during the test phase when creating their Post-test password. 
\item \textbf{Quiz compliance}. Whether participants correctly answered quiz questions tied to viewed tips.
\end{itemize}

\noindent Both constructs evaluate how participants respond to the rules they encounter.
To capture effects at different analytic resolutions, we compute at the \textbf{group-}, \textbf{user-}, and \textbf{rule-level}.

The resulting measures differ in their unit of analysis and in whether they apply to tip-linked behavior, quiz performance, or both.

\begin{itemize}
\item \textbf{Group-level tip compliance}. \emph{Unit of analysis = tip.} 
The proportion of all triggered test phase tips whose associated rules were satisfied in the post-test password, summarized per group.
\item \textbf{User-level tip compliance}. \emph{Unit of analysis = participant.}
The proportion of participants who satisfied all rules in their test phase rule set during post-test password creation.
\item \textbf{Rule-level tip compliance}. \emph{Unit of analysis = rule.} For each rule category, the proportion of participants who triggered that rule in the test phase and later satisfied it during post-test password creation.
\item \textbf{Rule-level quiz compliance}. \emph{Unit of analysis = rule.}
Accuracy on quiz questions targeting each rule category. This is evaluated using questions matched to test phase rules.
\item \textbf{User-level quiz compliance}. \emph{Unit of analysis = participant.}
For guided groups, each participant’s accuracy on questions whose target rules matched rules triggered in the test phase.
\end{itemize}

\subsubsection{Behavior-knowledge alignment} 
We define behavior-knowledge alignment as the proportion of triggered tips/associated rules that participants \emph{both} resolved during their post-test password creation and answered correctly on the exit quiz.

For measures where each participant contributes a single outcome (user-level and rule-level tip compliance), we report Wilson 95\% CI. For measures aggregated over multiple observations per participant (e.g., rule-level quiz compliance), we report cluster-robust 95\% CI obtained by nonparametric bootstrapping of users (4{,}000 resamples).

\section{Results}
\label{sec:results}

\noindent \emph{Can pedagogical friction support in situ learning and understanding of best practices for password generation?}

We evaluate in situ learning as compliance with the password rules/tips introduced in the test phase, measured in three ways during the post-test phase: tip compliance during password creation; quiz compliance on questions matched to those rules; and behavior-knowledge alignment comparing these two outcomes. Throughout, compliance is defined with respect to the rule categories in Table~\ref{tab:rule_set} and target rule categories in Appendix Table~\ref{tab:survey_questions}.

\subsection{Rule Reconstruction and Analytic Sample}
\label{subsubsec:tip_reconstruction}

\subsubsection{Rule reconstruction}
During the test phase, tips in guided conditions (T1, T2, T3) were triggered by associated rule violations and recorded. 
For quiz analyses that compare guided conditions with the control (T0), we derive a matched T0 baseline from its intermediate password snapshots. Specifically, because participants in T0 were not shown tips, the tips they \emph{would have been shown} were reconstructed from intermediate password keystroke logs. 
Measurement terms are defined in \S\ref{subsec:instrumentation_measures}.

\subsubsection{Analytic sample}
Our primary outcomes are rule-linked, but not all of them use the same analytic sample. Tip compliance and behavior-knowledge alignment require a set of rules that were explicitly introduced through tips in the test phase, so these analyses are restricted to the guided conditions T1–T3. The control condition T0 used an interface without instructional tips and therefore does not contribute to these tip compliance outcomes.

In quiz analyses, we examine questions matched to the corresponding rules. We also conduct comparisons by rule between T0 and T1–T3. This separation allows T0 to serve as a control baseline for quiz knowledge while preserving the logic of the tip compliance and alignment measures.

\subsection{Tip Compliance during Post-Test}
\label{subsubsec:tip_Compliance_in_Part2}

\subsubsection{Group-level tip compliance}
Fig.~\ref{fig:tip_compliance}(A) and Appendix Table~\ref{tab:tip_compliance_bygroup} show that tip compliance rates were high in all guided conditions: T1 91.9\%, T2 91.8\%, and T3 93.9\%. All three guidance designs led most tipped rules to be corrected in the post-test password. Unless otherwise noted, cluster-robust 95\% CIs were obtained by nonparametric bootstrapping of users, and Wilson 95\% CIs were used for single outcome proportions.

\subsubsection{User-level tip compliance}
Fig.~\ref{fig:tip_compliance}(B) and Appendix Table~\ref{tab:alltips_resolved_bygroup} show high all-rules-resolved rates across guided conditions: T1 81.8\%, T2 79.3\%, and T3 93.3\%. These values indicate that most participants who received tips in each group later complied with every rule they had been shown, with T3 showing the highest user-level compliance.

\subsubsection{Rule-level tip compliance}
\label{subsubsec:tip_compliance_by_rule}
Fig.~\ref{fig:tip_compliance}(C) shows that compliance was high for \textit{length}, \textit{uppercase}, \textit{number}, and \textit{personal information}, typically between 85\% and 100\% across groups (see Appendix Table~\ref{tab:per_rule_compliance} for details). Compliance for \textit{symbol} was lower and more variable across groups (e.g., 85.7\% in T3). 

\subsection{Quiz Compliance during Post-Test}
\subsubsection{Rule-level quiz compliance}
\label{subsubsec:rule_level_survey_compliance}

\begin{figure}[t]
  \centering
  \includegraphics[width=0.9\linewidth]{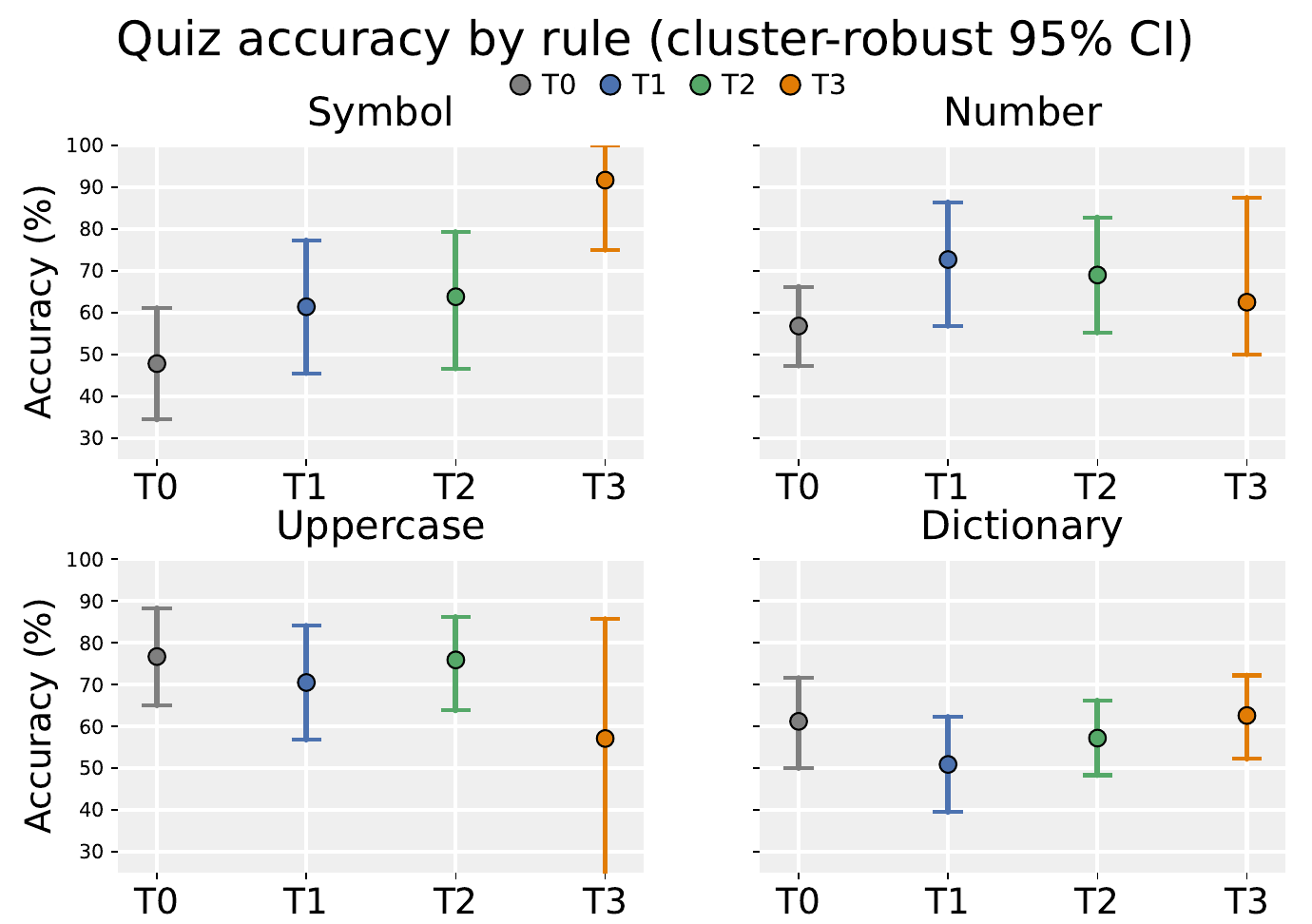}
  \caption{Quiz accuracy by rule for all conditions.}
  \label{fig:survey_compliance}
\end{figure}

We compare quiz accuracy by rule between a matched T0 baseline and participants in T1-T3 who had encountered that rule/associated tip. The largest descriptive differences appear for \textit{symbol} and \textit{number}. For \textit{symbol}, matched T0 accuracy is 47.8\%, compared with 61.4\% in T1, 63.8\% in T2, and 91.7\% in T3. For \textit{number}, matched T0 is 56.8\%, compared with 72.7\% in T1, 69.0\% in T2, and 62.5\% in T3. \textit{Uppercase} in T1-T3 showed little separation from T0, and \textit{dictionary} showed a mixed pattern. Personal information comparisons were not estimable under the snapshot reconstruction and are therefore omitted from this comparison. Fig.~\ref{fig:survey_compliance} and Appendix Table~\ref{tab:survey_rule_matched_t0} report the corresponding values and CIs.

Pairwise permutation tests identified a significant difference for \textit{symbol} between T3 and the T0 baseline after Holm correction ($p_{\mathrm{Holm}}=0.0003$), with T3 showing substantially higher quiz accuracy (91.7\% vs. 47.8\%). The remaining \textit{symbol} and \textit{number} comparisons were not statistically significant after correction (Appendix Table~\ref{tab:survey_inferential}). These results suggest that interactive guidance may be particularly helpful for symbol-related knowledge. 

\subsubsection{User-level quiz compliance}
We next summarize user-level quiz performance within the guided conditions. Participants typically saw several matched questions, one for each rule/tip from the test phase. 
We summarize user-level quiz performance by the proportion of matched questions each participant answered correctly.

Appendix Table~\ref{tab:survey_compliance_user} shows that mean accuracy is 62.8\% for T1, 63.4\% for T2, and 64.1\% for T3, based on 20, 23, and 24 users respectively. These user counts reflect the subset of participants in each group who both received tips and had at least one quiz question mapped to those rules, not the full group sizes. Our results suggest that participants in T1-T3 answered a similar proportion of their matched rule questions correctly.

\subsubsection{Behavior-Knowledge Alignment}
\label{subsubsec:bka}
\begin{figure}[t]
  \centering
  \includegraphics[width=\linewidth]{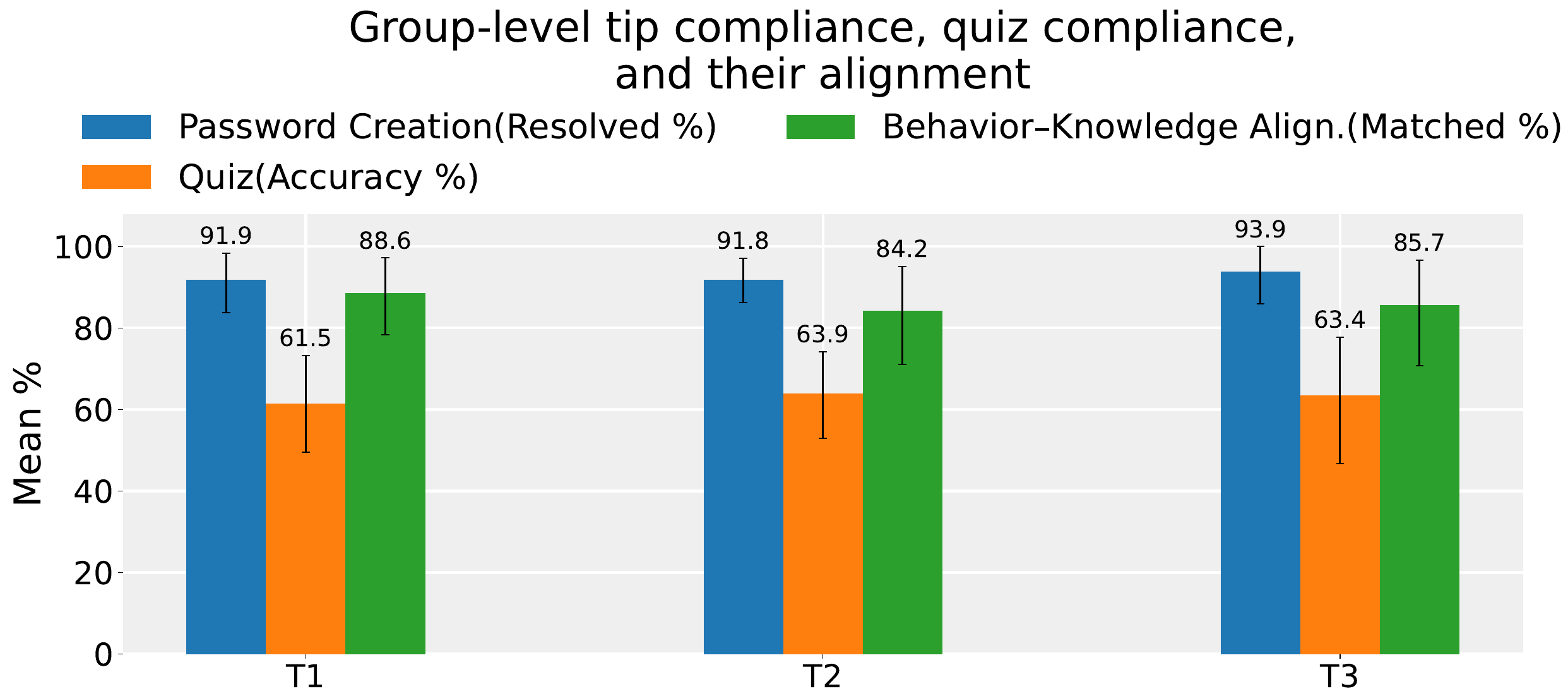}
  \caption{Group-level tip compliance in post-test phase, quiz compliance, and their alignment.}
  \label{fig:bka_by_group}
\end{figure}

Fig.~\ref{fig:bka_by_group} shows that behavior-knowledge alignment is high and similar across groups: 88.6\% for T1, 84.2\% for T2, and 85.7\% for T3. This illustrates that when participants resolved a rule in their post-test password, they usually answered at least one quiz question about that rule correctly.

\begin{figure*}[t]
  \centering
  \includegraphics[width=0.9\linewidth]{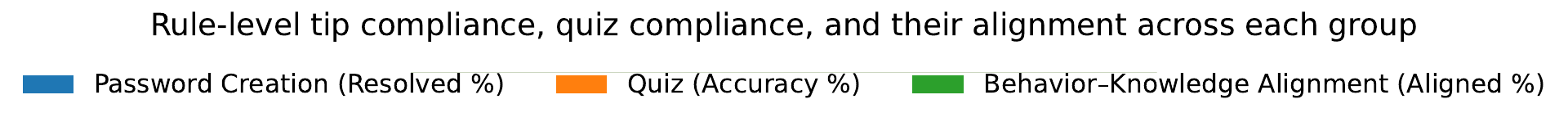}\\
  \begin{minipage}{0.3\textwidth}\centering
    \includegraphics[width=\linewidth,height=3.5cm]{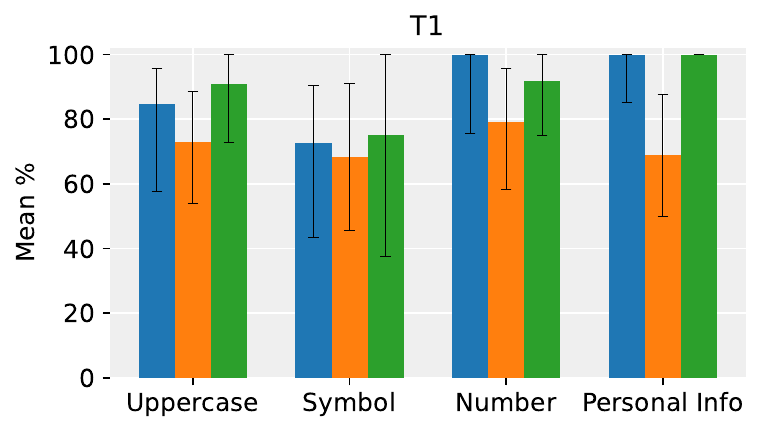}
  \end{minipage}
  \begin{minipage}{0.3\textwidth}\centering
    \includegraphics[width=\linewidth,height=3.5cm]{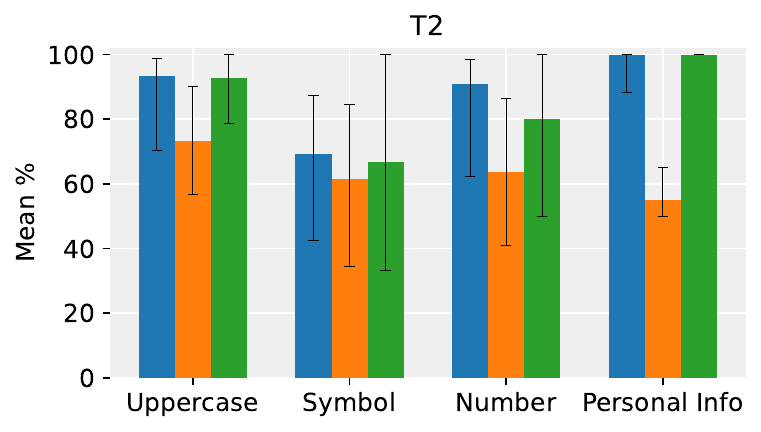}
  \end{minipage}
  \begin{minipage}{0.3\textwidth}\centering
    \includegraphics[width=\linewidth,height=3.5cm]{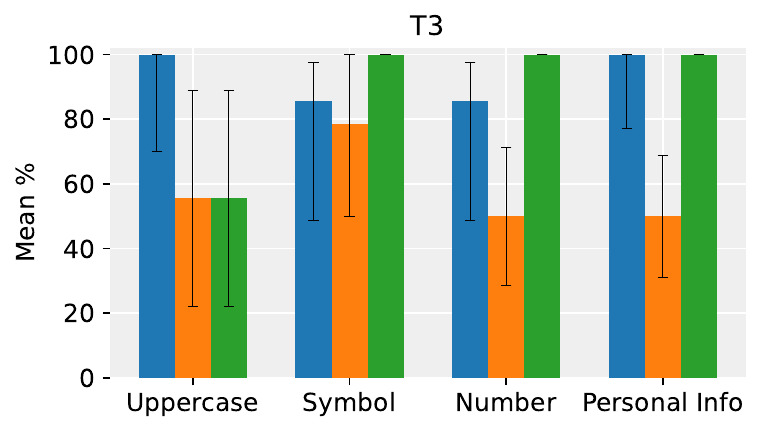}
  \end{minipage}
  \caption{Rule-level tip compliance, quiz compliance, and their alignment across each group.}
  \label{fig:bka_per_rule_by_group}
\end{figure*}

At the rule level, the quiz did not include any questions whose target rule was \textit{length}, and \textit{dictionary} was not part of the Post-test rule checks, so these two rules are omitted. Alignment for the remaining rules is generally high (Appendix Fig.~\ref{fig:bka_per_rule_by_group} and Table~\ref{tab:bka_rule}). \textit{Symbol} and \textit{number} have their highest alignment in T3 (around 100\%). \textit{Personal information} is at ceiling in all three groups. These rule-level patterns are consistent with the idea in \S\ref{subsubsec:rule_level_survey_compliance} that the interactive friction in T3 is especially helpful for remembering concrete rules such as \textit{symbol} and \textit{number} once they have been applied in the post-test password creation.

To understand how behavior and knowledge diverged, we also examined two kinds of misalignment, \emph{behavior-knowledge misalignment} and \emph{knowledge-behavior misalignment}. Further details are provided in Appendix~\ref{appendix_misalign}.

\subsection{Digital Literacy}

\noindent We examine whether age, gender, education, and digital literacy were associated with the compliance outcomes. Digital literacy was measured using the ten-item instrument of van Deursen et al.~\cite{van2014measuring}. Results are provided in Appendix~\ref{appendix_demo_dig}.

\section{Discussion}
\subsection{Compliance}
\noindent We recorded strength scores provided by the Password Guessability Service \cite{pgs-service} but do not treat them as a primary outcome. Compliance aligns directly with our manipulation, since it evaluates whether participants resolved the specific rules introduced by the tips (learning). In contrast, the strength score is a single composite measure that does not provide the rule-linked instruction or active acknowledgment, and cannot capture how closely behavior aligns with the rules targeted by our intervention. 

Two design properties of pedagogical friction may help explain observed high compliance: immediate, rule-linked feedback at the moment of editing; and, an interactive acknowledgment step in T3 that encouraged attention to guidance. Both properties are consistent with microlearning principles and behavioral models that emphasize concise, in-flow guidance and timely prompts~\cite{hug2006microlearning,allela2021introduction,sankaranarayanan2023microlearning,DBLP:journals/ubiquity/Fogg02}. Throughout the remainder of this subsection, \emph{compliance} refers to group-level tip compliance.

\noindent \textbf{Immediate and rule-linked feedback.}
Our analyses use the set of rules participants were shown during the test phase (each item of guidance is linked to a named rule in Table~\ref{tab:rule_set}). Consistent with this interpretation, participants tended to resolve the mapped rules in their post-test phase passwords (Fig.~\ref{fig:tip_compliance}(A); Appendix Table~\ref{tab:tip_compliance_bygroup}), and many resolved all mapped rules (Fig.~\ref{fig:tip_compliance}(B)). 

Our pedagogical frictions recommended a minimal change and gave the user direct control over how to implement that change. This pairing of a concrete instruction with an immediate opportunity to act supports psychological capability and opportunity, e.g., \cite{michie2011behaviour,petty2012communication}. Rule-level analyses corroborate this view: when instructions were simple to operationalize, such as \textit{length} or \textit{uppercase}, compliance often approached or reached ceiling, whereas \textit{symbol} showed more residual misses, consistent with a higher search cost or uncertainty about which characters qualify (Fig.~\ref{fig:tip_compliance}(C); Table~\ref{tab:per_rule_compliance}).

\noindent \textbf{Interactive acknowledgment.}
In T3, users briefly acknowledged guidance before proceeding. This light-touch action serves two roles. First, it ensures attention. Acknowledgment requires a momentary shift from scanning to recognition, which can reduce the chance that a tip is ignored. Second, the acknowledgment click constitutes a small active \textit{enactment} step. Evidence from memory research shows that performing an action during encoding strengthens later memory, particularly in brief learning episodes~\cite{roediger2006power}. These mechanisms are consistent with the higher all-rules-resolved rates in T3 (Fig.~\ref{fig:tip_compliance}(B)).

\subsection{Behavior and Knowledge Alignment}

\noindent Our alignment measure focuses on how often people who followed a rule when creating their post-test password also demonstrated that they knew that rule on the quiz. Misalignment measures complement this by showing how often mismatches lean in each direction, with behavioral compliance more common than explicit rule recognition. I.e., behavior-knowledge misalignment was more frequent, wherein people changed their passwords to follow tips they received in the test phase but did not pick the correct answer to the matched question on the quiz. The opposite pattern, knowledge-behavior misalignment, was rare. I.e., it was uncommon for people to choose the correct response on the exit quiz but ignore the associated rule during their post-test password creation. In brief, learning more often appears to move from action to rule than from rule to action.

In settings where explicit understanding is important, designers may want to add small follow-up cues that link a successful edit back to the rule description, instead of relying only on a one-time rule recall after the task.

\subsection{Practical Implications for Password Creation}

\noindent Our work suggests that, for password creation, brief embedded guidance can change the actions users take and may leave them with short-term memory of some of the rules they encountered. At the same time, weak passwords do not arise only from missing knowledge. Password behavior is also shaped by memory demands, time pressure, competing goals, and the broader burden of managing many accounts. Thus, we present pedagogical friction as a targeted interface intervention within these constraints, not as a complete solution to password security.

We observed high correction of rule-linked issues across the three guided conditions, so our recommendations focus on how to select and implement guidance styles that improve password quality and help protect accounts that hold sensitive personal information.

\noindent \textbf{Choose the lightest effective intervention for the context.} 
Start with minimal guidance that is likely to work in the target setting. For familiar, lower risk actions, concise prompts that name the violated rule and the required fix are often sufficient. For higher assurance scenarios, richer guidance or a short interactive step may be justified.

\noindent \textbf{Brief guidance.} 
Use short, specific messages when speed and low burden are priorities. Tie each prompt to a single rule and present one prompt at a time. Provide a clear success indicator once the requirement is met.

\noindent \textbf{Detailed guidance.} 
Use compact explanations when the goal includes short-term understanding of selected rules in addition to correction. Pair the explanation with the concrete instruction, and allow users to reopen the same guidance within the session without leaving the task.

\noindent \textbf{Interactive acknowledgment guidance.} 
Require active engagement when completeness at the point of creation is essential. Because such engagement adds effort, use it sparingly. A staged strategy can be useful: require acknowledgment after repeated violations, and relax it once consistent compliance is observed.

\noindent \textbf{Address rule-specific and validator transparency issues.}
Rule-level results indicate that some rules were easier to satisfy and recognize than others. In our findings, \textit{symbol} remained the most variable rule in behavioral compliance, while the quiz analysis showed the clearest benefit for symbol-related knowledge: T3 significantly outperformed T0 after Holm correction. To improve consistency, make acceptance rules visible at the point of entry: show a brief example of acceptable characters, clarify any restrictions, and confirm acceptance immediately after the change. When assessing knowledge with quiz questions, avoid multi-rule quiz questions.

Potential applications beyond password creation, including privacy consent and permission settings, are discussed in Appendix~\ref{appendix_broader_applications}.

\subsection{Limitations and Future Work}

\noindent One limitation of this work concerns the scope of our task and setting. We study a single password creation task on a custom website as one application of pedagogical friction, using a fixed rule set and a moderate level of perceived risk. This design lets us examine how pedagogical friction affects rule compliance and short-term quiz performance in a controlled environment, but it does not show that the same patterns would hold for higher- or lower-stakes accounts, which may elicit different levels of attention, motivation, and tolerance for interruptions. Future work should test pedagogical friction in other settings and decision tasks.

A second limitation is the short retention interval. We were interested in whether participants would remember and reuse what they learned after the initial test phase, but in practice the post-test phase occurred directly after the test phase to keep participants engaged and limit dropout. Our results therefore speak only to short term recall and near term transfer. Future studies should adopt longer delays and follow-up sessions that better approximate real world use, for example multi-day or multi-week intervals, to test whether pedagogical friction supports sustained changes in knowledge retention and behavior-knowledge alignment.

A third limitation concerns the design of the intervention itself. All four conditions included the same password strength meter, which allowed us to study the added role of pedagogical tips beyond a common baseline interface, but reduced our ability to attribute changes in overall password strength scores specifically to the tips displayed. In addition, the interface could trigger up to 20 possible tips, and participants saw different subsets depending on the passwords they entered. This introduces heterogeneity in exposure: observed differences may reflect both friction level and the specific mix of rules each participant encountered. Our rule-level analyses partially address this issue, but they do not remove it entirely. For these reasons, our strongest evidence comes from rule-linked compliance and quiz analyses rather than from the strength score.

\section{Ethics Statement}
\noindent Our study protocol was reviewed and approved by our IRB before recruitment. Participants provided informed consent and could stop at any time without penalty. The study posed minimal risk and involved a password creation task with the interface variations described in the paper.

To protect privacy, we linked study data only to an anonymous participant identifiers. Our platform stored password entries and keystrokes on a secured, access controlled server so we could compute password related measures. Raw password strings and keystroke logs were not included in the analytic dataset and were not shared outside the research team. Any shared results and materials use de-identified, derived measures.

Participants were paid US\$15 per hour, prorated for time spent. The study used mild deception in the form of a brief cover story about the study purpose to preserve study validity. This deception posed minimal risk and did not withhold information needed for consent about procedures, data collection, or compensation. Participants were debriefed at the end of the study, and we did not collect participant images, audio, or video.

\section{Conclusion}

\noindent We have introduced pedagogical friction to bring short, rule-linked explanations into the process of password creation, and evaluated three frictional designs in a controlled web environment. We found that pedagogical friction supported compliance with password rules and high behavior-knowledge alignment. Quiz accuracy was significantly higher for T3 than T0 on symbol-related questions after correction. We provide an experimental template, measures, and analysis scripts that others can reuse to test when and how carefully calibrated friction supports users’ responses to digital guidance and more informed actions online.

\bibliographystyle{IEEEtranN}
\bibliography{ref}

\appendices

\section{Demographics}

Participant demographics are summarized in Table~\ref{tab:demographics}. 

\begin{table}[t]
\small
\setlength{\tabcolsep}{3pt}
\renewcommand{\arraystretch}{1.05}
\centering
\caption{Participant demographics summary.}
\label{tab:demographics}
\begin{tabular}{@{}p{3.0cm}r|p{3cm}r@{}}
\toprule
\textbf{Gender} & \% & \textbf{Age} & \% \\
\midrule
Female & 53.1 & 18--24 & 14.1 \\
Male & 43.8 & 25--34 & 35.9 \\
Non-binary/Third gender & 1.6 & 35--44 & 21.9 \\
Prefer not to say & 1.6 & 45--54 & 14.1 \\
 &  & 55--64 & 9.4 \\
 &  & 65+ & 4.7 \\
\midrule
\textbf{Ethnicity} & \% & \textbf{Education} & \% \\
\midrule
White or Caucasian & 61.7 & Graduate or Professional degree & 18.8 \\
Black or African American & 21.9 & Bachelor’s degree & 42.2 \\
Asian & 8.6 & Associates or Technical degree & 10.2 \\
Other & 3.1 & Some college, no degree & 18.8 \\
American Indian/Alaska Native & 0.8 & High school diploma or GED & 9.4 \\
Mixed/Multiple (e.g., White + Asian, etc.) & 3.2 & Prefer not to say & 0.8 \\
Prefer not to say & 0.8 &  &  \\
\bottomrule
\end{tabular}
\end{table}

\section{Supplementary details for compliance and alignment results}
\label{appendix_compliance}

This section reports detailed numerical values for the compliance and alignment analyses summarized in \S\ref{sec:results}.

Table~\ref{tab:tip_compliance_bygroup} expands the group-level tip compliance results reported in Fig.~\ref{fig:tip_compliance}(A). It shows how many tips of each guided group were triggered in the test phase (\textit{Tips $n$}), how many of those tips were resolved in the post-test password (\textit{Resolved $n$}), the corresponding percentage. All three guided conditions have high resolution rates around 92-94\%.

Table~\ref{tab:per_rule_compliance} reports rule-level tip compliance. Because no dictionary reference list was available for analysis, the \textit{dictionary} rule was omitted from this evaluation. The \textit{Users tipped} column reports how many participants were shown the tip for that rule in the test phase. Results are presented in \S\ref{sec:results}.

\begin{table}[t]
\centering
\caption{Group-level tip compliance (cluster-robust 95\% CI).}
\label{tab:tip_compliance_bygroup}
\fontsize{8}{9.5}\selectfont
\begin{tabular}{
    >{\centering\arraybackslash}p{1cm} |
    >{\centering\arraybackslash}p{0.8cm}
    >{\centering\arraybackslash}p{1.4cm}
    >{\centering\arraybackslash}p{1.4cm}
    >{\centering\arraybackslash}p{1.8cm}}
\toprule
\textbf{Group} & \textbf{Tips $n$} & \textbf{Resolved $n$} & \textbf{Resolved \%} & \textbf{95\% CI} \\
\midrule
T1 & 62 & 57 & 91.9 & [83.8--98.4] \\
T2 & 73 & 67 & 91.8 & [86.3--97.1] \\
T3 & 49 & 46 & 93.9 & [86.0--100] \\
\bottomrule
\end{tabular}
\end{table}

\begin{table}[t]
    \centering  
    { 
        \caption{Rule-level tip compliance (Wilson 95\% CI).}
        \label{tab:per_rule_compliance}
        \fontsize{7}{8.5}\selectfont
        \begin{tabular}{
            >{\centering\arraybackslash}p{1.0cm}|
            >{\centering\arraybackslash}p{0.7cm}
            >{\centering\arraybackslash}p{1.0cm}
            >{\centering\arraybackslash}p{0.8cm}
            >{\centering\arraybackslash}p{0.8cm}
            >{\centering\arraybackslash}p{1.5cm}}
        \toprule
        \textbf{Rule} & \textbf{Group} & \textbf{Users tipped $n$} & \textbf{Resolved $n$} & \textbf{Resolved \%} & \textbf{95\% CI} \\
        \midrule
        \multirow{3}{*}{Length} &
        T1 & 22 & 22 & 100 & [85.1--100] \\
        & T2 & 29 & 29 & 100 & [88.3--100] \\
        & T3 & 13 & 13 & 100 & [77.2--100] \\
        \midrule
        \multirow{3}{*}{Symbol} &
        T1 & 11 &  8 &  72.7 & [43.4--90.3] \\
        & T2 & 13 &  9 &  69.2 & [42.4--87.3] \\
        & T3 &  7 &  6 &  85.7 & [48.7--97.4] \\
        \midrule
        \multirow{3}{*}{Uppercase} &
        T1 & 13 & 11 &  84.6 & [57.8--95.7] \\
        & T2 & 15 & 14 &  93.3 & [70.2--98.8] \\
        & T3 &  9 &  9 & 100 & [70.1--100] \\
        \midrule
        \multirow{3}{*}{Number} &
        T1 & 12 & 12 & 100 & [75.7--100] \\
        & T2 & 11 & 10 &  90.9 & [62.3--98.4] \\
        & T3 &  7 &  6 &  85.7 & [48.7--97.4] \\
        \midrule
        \multirow{3}{*}{\makecell{Personal\\Info}} &
        T1 &  4 &  4 & 100 & [51.0--100] \\
        & T2 &  5 &  5 & 100 & [56.6--100] \\
        & T3 &  4 &  4 & 100 & [51.0--100] \\
        \bottomrule
        \end{tabular}
    }   
\end{table}

Table~\ref{tab:alltips_resolved_bygroup} summarizes user-level tip compliance. For each guided group, it reports how many participants who received tips in the test phase resolved all of their triggered rules in the post-test password (\textit{All Resolved $n$}), the corresponding percentage. In all three guided conditions, a large majority of tipped users resolved every rule that had triggered a tip, with T1 and T2 around 80\% and T3 above 90\%.

\begin{table}[t]
\centering
\caption{User-level tip compliance (Wilson 95\% CI).}
\label{tab:alltips_resolved_bygroup}
\fontsize{8}{9.5}\selectfont
\begin{tabular}{
    >{\centering\arraybackslash}p{1cm} |
    >{\centering\arraybackslash}p{2.0cm}
    >{\centering\arraybackslash}p{2.0cm}
    >{\centering\arraybackslash}p{1.8cm}}
\toprule
\textbf{Group} & \textbf{All Resolved $n$} & \textbf{All Resolved \%} & \textbf{95\% CI} \\
\midrule
T1 & 18 & 81.8 & [61.5--92.7] \\
T2 & 23 & 79.3 & [61.6--90.2] \\
T3 & 28 & 93.3 & [78.7--98.2] \\
\bottomrule
\end{tabular}
\end{table}

Table~\ref{tab:survey_compliance_user} summarizes user-level quiz compliance within the guided conditions. It reports the number of participants who both received tips and had at least one matched quiz question (\textit{Users $n$}) and their mean accuracy on those questions. Mean accuracy is just above 60\% in all three guided conditions, showing similar user-level performance across T1-T3.

\begin{table}[t]
\centering
\caption{User-level quiz compliance within guided conditions (cluster-robust 95\% CI).}
\label{tab:survey_compliance_user}
\fontsize{8}{9.5}\selectfont
\begin{tabular}{
    >{\centering\arraybackslash}p{0.8cm} |
    >{\centering\arraybackslash}p{1.5cm}
    >{\centering\arraybackslash}p{2.1cm}
    >{\centering\arraybackslash}p{1.6cm}}
\toprule
\textbf{Group} & \textbf{Users $n$} & \textbf{Mean accuracy \%} & \textbf{95\% CI} \\
\midrule
T1 & 20 & 62.8 & [52.4--72.5] \\
T2 & 23 & 63.4 & [52.8--73.8] \\
T3 & 24 & 64.1 & [50.3--76.2] \\
\bottomrule
\end{tabular}
\end{table}

\begin{table}[t]
\centering
\caption{Behavior and knowledge misalignment by group.}
\label{tab:bk_misalign_group}
\fontsize{8}{9.5}\selectfont
\begin{tabular}{
    >{\centering\arraybackslash}p{1.4cm} |
    >{\centering\arraybackslash}p{2.8cm}
    >{\centering\arraybackslash}p{2.8cm}}
\toprule
\textbf{Group} & \textbf{Behavior-knowledge (\% of cases)} & \textbf{Knowledge-behavior (\% of cases)} \\
\midrule
T1 & 40.0\% &  7.5\% \\
T2 & 45.5\% &  4.5\% \\
T3 & 48.4\% &  3.2\% \\
\bottomrule
\end{tabular}
\end{table}

Table~\ref{tab:survey_inferential} presents the results of pairwise permutation comparisons of quiz accuracy for the two rule categories that showed the largest apparent differences: \textit{Symbol} and \textit{Number}. For each rule, we compared T1, T2, and T3 against the T0 baseline and applied Holm correction across the six tested contrasts. After Holm correction, the T3 versus T0 comparison for Symbol was statistically significant ($p_{\mathrm{Holm}}=0.0003$); the remaining five comparisons were not statistically significant.

\begin{table}[t]
\centering
\caption{Pairwise comparisons for quiz accuracy.}
\label{tab:survey_inferential}
\fontsize{8}{9.5}\selectfont
\begin{tabular}{
    >{\centering\arraybackslash}p{1.0cm} |
    >{\centering\arraybackslash}p{1.2cm}
    >{\centering\arraybackslash}p{1.0cm}
    >{\centering\arraybackslash}p{0.9cm}
    >{\centering\arraybackslash}p{1.0cm}
    >{\centering\arraybackslash}p{0.9cm}}
\toprule
\textbf{Rule} & \textbf{Contrast} & \textbf{Treatment mean \%} & \textbf{T0 mean \%} & \textbf{Difference} & \textbf{$p_{\mathrm{Holm}}$} \\
\midrule
Symbol & T1 vs T0 & 61.4 & 47.8 & +13.6 & 0.651 \\
Symbol & T2 vs T0 & 63.8 & 47.8 & +16.0 & 0.651 \\
Symbol & T3 vs T0 & 91.7 & 47.8 & +43.9 & 0.0003 \\
Number & T1 vs T0 & 72.7 & 56.8 & +16.0 & 0.483 \\
Number & T2 vs T0 & 69.0 & 56.8 & +12.2 & 0.651 \\
Number & T3 vs T0 & 62.5 & 56.8 & +5.7 &  0.651 \\
\bottomrule
\end{tabular}
\end{table}

Table~\ref{tab:bka_rule} breaks down behavior-knowledge alignment by rule. For each rule and group, it reports how many participants had at least one resolved instance of that rule in the post-test password and at least one matched quiz question (\textit{Users $n$}), and the mean alignment percentage. Alignment is generally high across rules, especially for \textit{Personal information} and for \textit{Number} and \textit{Symbol} in T3. Estimates for some rules, such as \textit{Personal information} and \textit{Symbol}, are based on relatively few users in each group, so the corresponding CIs are wider.

\begin{table}[t]
\centering
\caption{Behavior and knowledge alignment by rule (cluster-robust 95\% CI).}
\label{tab:bka_rule}
\fontsize{8}{9.5}\selectfont
\begin{tabular}{
    >{\centering\arraybackslash}p{1.3cm} |
    >{\centering\arraybackslash}p{1.0cm}
    >{\centering\arraybackslash}p{1.0cm}
    >{\centering\arraybackslash}p{1.0cm}
    >{\centering\arraybackslash}p{1.9cm}}
\toprule
\textbf{Rule} & \textbf{Group} & \textbf{Users $n$} & \textbf{Mean \%} & \textbf{95\% CI} \\
\midrule
\multirow{3}{*}{Uppercase} &
T1 & 11 & 90.9 & [72.7--100] \\
& T2 & 14 & 92.9 & [78.6--100] \\
& T3 &  9 & 55.6 & [22.2--88.9] \\
\midrule
\multirow{3}{*}{Symbol} &
T1 &  8 & 75.0 & [37.5--100] \\
& T2 &  9 & 66.7 & [33.3--100] \\
& T3 &  8 & 100 & [100--100] \\
\midrule
\multirow{3}{*}{Number} &
T1 & 12 & 91.7 & [75.0--100] \\
& T2 & 10 & 80.0 & [50.0--100] \\
& T3 &  7 & 100 & [100--100] \\
\midrule
\multirow{3}{*}{\makecell{Personal\\Info}} &
T1 &  4 & 100 & [100--100] \\
& T2 &  5 & 100 & [100--100] \\
& T3 &  4 & 100 & [100--100] \\
\bottomrule
\end{tabular}
\end{table}

\subsection{Divergence in Behavior and Knowledge}
\label{appendix_misalign}

To understand how behavior and knowledge diverged, we examined two kinds of misalignment. \textbf{Behavior-knowledge misalignment} captures cases where a rule was resolved in the post-test password but none of the matched quiz questions about that rule were answered correctly. \textbf{Knowledge-behavior misalignment} captures the complementary pattern in which the rule was not resolved in the post-test password even though the matched questions were answered correctly.

Table~\ref{tab:bk_misalign_group} summarizes these misalignments as proportions of all rule-level cases with both post-test behavior and matched quiz data. Behavior-knowledge misalignment occurs for a minority of cases in each group, while knowledge-behavior misalignment is rare. Together with the high alignment rates in \S\ref{subsubsec:bka}, this pattern indicates that when participants applied a rule in their post-test password, they usually answered the quiz question about that rule correctly, and situations where they appeared to ``know'' a rule but did not use it in their post-test password were uncommon.

\section{Digital Literacy Results}
\label{appendix_demo_dig}
Digital literacy was measured using the ten statements from van Deursen et al.'s digital skills instrument~\cite{van2014measuring} (Table~\ref{tab:digital_literacy}), each rated on a 5-point scale. We combined responses into a single literacy score and grouped participants into quintiles for analysis, higher quintiles indicating greater digital literacy. We then examined whether age, gender, education, and digital literacy were associated with the compliance outcomes. In these analyses, demographic levels labeled \textit{Prefer not to say} were omitted because they contained too few participants for stable estimates.

For group-level compliance, rates remained high in all demographic groups and across the full digital literacy range, with most strata above 85\% (Table~\ref{tab:tip_compliance_demo_lit}). At the user level, all-rules-resolved rates showed a similar pattern, and the apparent fluctuations in some categories were driven by very small subgroups (e.g., the 65+ age, which included only 1-2 users; Table~\ref{tab:all_rules_resolved_demo_lit}).

With respect to quiz compliance, accuracy was also similar across education levels, genders, age (Table~\ref{tab:survey_demo_lit}). Most strata fall in the range of roughly 60-75\% correct. Behavior-knowledge alignment showed the same pattern.

These analyses show that compliance during password creation, quiz understanding, and behavior-knowledge alignment were consistently comparable across age, gender, education, and literacy groups, with no evidence that any subgroup was disadvantaged. The intervention’s effects therefore appear robust and broadly accessible, supporting similar learning and recall for all participants.

Table~\ref{tab:rule_mappping} lists the mapping between core rules and quiz target rules.

Table~\ref{tab:survey_questions} lists the quiz questions and its 20 target rules. The \textit{Target Rule} is for analysis and reporting only. Participants saw only the questions and question options.

\begin{table}[t]
\centering
\caption{Group-level tip compliance by demographic and digital literacy factors.}
\label{tab:tip_compliance_demo_lit}
\fontsize{8}{9.5}\selectfont
\begin{tabular}{
    >{\centering\arraybackslash}p{1.3cm} |
    >{\centering\arraybackslash}p{2.8cm}
    >{\centering\arraybackslash}p{0.8cm}
    >{\centering\arraybackslash}p{0.8cm}
    >{\centering\arraybackslash}p{0.8cm}}
\toprule
\textbf{Factor} & \textbf{Characteristic} & \textbf{T1} & \textbf{T2} & \textbf{T3} \\
\midrule
\multirow{6}{*}{Education}
& High school / GED                  & 100 & 100 & 100 \\
& Some college                       &  91.7 &  90.9 & 100 \\
& Associates / technical             & 100 &  83.3 & 100 \\
& Bachelor’s degree                  &  94.1 &  96.4 &  87.0 \\
& Graduate / professional            &  87.5 &  86.7 & 100 \\
\midrule
\multirow{3}{*}{Gender}
& Female                             &  92.3 &  91.7 &  89.3 \\
& Male                               &  94.1 &  92.0 & 100 \\
\midrule
\multirow{6}{*}{Age}
& 18--24                             & 100 &  85.7 & 100 \\
& 25--34                             & 100 &  96.6 &  95.0 \\
& 35--44                             & 100 &  87.0 & 100 \\
& 45--54                             &  84.6 & 100 & 100 \\
& 55--64                             &  77.8 & 100 & 100 \\
& 65+                                &  66.7 &  66.7 &  60.0 \\
\bottomrule
\end{tabular}
\end{table}

\begin{table}[t]
\centering
\caption{User-level tip compliance by demographic and digital literacy factors.}
\label{tab:all_rules_resolved_demo_lit}
\fontsize{8}{9.5}\selectfont
\begin{tabular}{
    >{\centering\arraybackslash}p{1.3cm} |
    >{\centering\arraybackslash}p{2.8cm}
    >{\centering\arraybackslash}p{0.8cm}
    >{\centering\arraybackslash}p{0.8cm}
    >{\centering\arraybackslash}p{0.8cm}}
\toprule
\textbf{Factor} & \textbf{Characteristic} & \textbf{T1} & \textbf{T2} & \textbf{T3} \\
\midrule
\multirow{6}{*}{Education}
& High school / GED                  & 100 & 100 & 100 \\
& Some college                       &  75.0 &  71.4 & 100 \\
& Associates / technical             & 100 &  66.7 & 100 \\
& Bachelor’s degree                  &  91.7 &  91.7 &  84.6 \\
& Graduate / professional            &  66.7 &  66.7 & 100 \\
\midrule
\multirow{3}{*}{Gender}
& Female                             &  75.0 &  77.8 &  88.2 \\
& Male                               &  92.3 &  81.8 & 100 \\
\midrule
\multirow{6}{*}{Age}
& 18--24                             & 100 &  66.7 & 100 \\
& 25--34                             & 100 &  90.0 &  92.3 \\
& 35--44                             & 100 &  62.5 & 100 \\
& 45--54                             &  50.0 & 100 & 100 \\
& 55--64                             &  66.7 & 100 & 100 \\
& 65+                                &   0 &  50.0 &   0 \\
\bottomrule
\end{tabular}
\end{table}

\begin{table}[t]
    \centering  
    {
        \caption{RULE-LEVEL QUIZ COMPLIANCE (CLUSTER-ROBUST 95\% CI).}
        \label{tab:survey_rule_matched_t0}
        \fontsize{7}{8.5}\selectfont
        \begin{tabular}{
            >{\centering\arraybackslash}p{1.1cm}|
            >{\centering\arraybackslash}p{0.75cm}
            >{\centering\arraybackslash}p{0.95cm}
            >{\centering\arraybackslash}p{0.8cm}
            >{\centering\arraybackslash}p{1.45cm}}
        \toprule
        \textbf{Rule} & \textbf{Group} & \textbf{Questions $n$} & \textbf{Accuracy \%} & \textbf{95\% CI} \\
        \midrule
        \multirow{4}{*}{Uppercase} &
        T0 & 60 & 76.7 & [65.0--88.3] \\
        & T1 & 44 & 70.5 & [56.8--84.1] \\
        & T2 & 58 & 75.9 & [63.8--86.2] \\
        & T3 & 14 & 57.1 & [14.3--85.7] \\
        \midrule
        \multirow{4}{*}{Symbol} &
        T0 & 90 & 47.8 & [34.4--61.1] \\
        & T1 & 44 & 61.4 & [45.5--77.3] \\
        & T2 & 58 & 63.8 & [46.6--79.3] \\
        & T3 & 12 & 91.7 & [75.0--100] \\
        \midrule
        \multirow{4}{*}{Number} &
        T0 & 74 & 56.8 & [47.3--66.2] \\
        & T1 & 44 & 72.7 & [56.8--86.4] \\
        & T2 & 58 & 69.0 & [55.2--82.8] \\
        & T3 & 8 & 62.5 & [50.0--87.5] \\
        \midrule
        \multirow{4}{*}{Dictionary} &
        T0 & 250 & 61.2 & [50.0--71.6] \\
        & T1 & 220 & 50.9 & [39.5--62.3] \\
        & T2 & 290 & 57.2 & [48.3--66.2] \\
        & T3 & 230 & 62.6 & [52.2--72.2] \\
        \bottomrule
        \end{tabular}
    }
\end{table}

\begin{table}[t]
\centering
\begin{threeparttable}
\caption{Quiz compliance by demographic and digital literacy factors.}
\label{tab:survey_demo_lit}
\fontsize{8}{9}\selectfont
\begin{tabular}{
    >{\centering\arraybackslash}p{1.5cm} |
    >{\centering\arraybackslash}p{3.0cm}
    >{\centering\arraybackslash}p{0.8cm}
    >{\centering\arraybackslash}p{1.6cm}}
\toprule
\textbf{Factor} & \textbf{Characteristic} & \textbf{Counts} & \textbf{Accuracy (\%)} \\
\midrule
\multirow{5}{*}{Education} 
& High school / GED & 4 & 62--79 \\
& Some college & 13 & 52--78 \\
& Associates / technical & 5 & 62--94 \\
& Bachelor’s degree & 30 & 47--70 \\
& Graduate / professional & 14 & 51--79 \\
\midrule
\multirow{2}{*}{Gender}
& Female & 36 & 54--71 \\
& Male & 30 & 54--75 \\
\midrule
\multirow{6}{*}{Age}
& 18--24 & 7 & 65--87 \\
& 25--34 & 27 & 50--73 \\
& 35--44 & 14 & 47--74 \\
& 45--54 & 9 & 47--79 \\
& 55--64 & 7 & 64--94 \\
& 65+ & 3 & 10--50 \\
\midrule
\multirow{5}{*}{\makecell{Digital\\Literacy}}
& Q1 (Lowest) & 16 & 59--81 \\
& Q2 & 12 & 44--77 \\
& Q3 & 14 & 58--84 \\
& Q4 & 15 & 53--80 \\
& Q5 (Highest) & 10 & 22--53 \\
\bottomrule
\end{tabular}
\begin{tablenotes}[para,flushleft]
\footnotesize
\textit{Note.} \textbf{Counts} is the number of users whose quiz responses included at least one question that matched the rules they were tipped on in the test phase.
\end{tablenotes}
\end{threeparttable}
\end{table}

\begin{table}[t]
\centering
\caption{Digital literacy statements.}
\label{tab:digital_literacy}
\fontsize{8.5}{10}\selectfont
\begin{tabular}{
    >{\centering\arraybackslash}p{1.0cm} |
    >{\raggedright\arraybackslash}p{6.3cm}}
\toprule
\textbf{Item} & {\centering\textbf{Statement}} \\
\midrule
1 & It is easy for me to find information online. \\
2 & I should take a course on finding information online. \\
3 & I know how to use a wide range of strategies when searching for information online. \\
4 & I find it hard to decide what the best keywords are to use for online searches. \\
5 & I am confident selecting results from a search engine. \\
6 & I normally look at more than the top three search results. \\
7 & Sometimes I find it hard to verify information I have retrieved. \\
8 & I feel confident in my evaluation of whether a website can be trusted. \\
9 & I generally compare different websites to decide if information is true. \\
10 & I carefully consider the information I find online. \\
\bottomrule
\end{tabular}
\end{table}

\begin{table}[t]
\centering
\caption{Mapping between core rules and quiz target rules.}
\label{tab:rule_mappping}
\fontsize{7.5}{9}\selectfont
\begin{tabular}{
    >{\centering\arraybackslash}m{1.5cm} |
    >{\raggedright\arraybackslash}m{5.8cm}}
\toprule
\textbf{Rule} & \textbf{Quiz target rule} \\
\midrule

\multirow{2}{*}{\textit{Symbol}} &
Add symbols in unpredictable locations. \\
\cmidrule{2-2}
 & Move symbols and digits elsewhere in your password. \\
\midrule

\multirow{2}{*}{\textit{Uppercase}} &
Capitalize a letter in the middle. \\
\cmidrule{2-2}
 & Mix up your capitalization. \\
\midrule

\multirow{2}{*}{\textit{Number}} &
Consider inserting digits into the middle. \\
\cmidrule{2-2}
 & Avoid numerical patterns. \\
\midrule

\multirow{5}{*}{\textit{Personal Info}} &
Don't use site-specific terms in your password. \\
\cmidrule{2-2}
 & Don't use names. \\
\cmidrule{2-2}
 & Don't use pet names. \\
\cmidrule{2-2}
 & Avoid using dates. \\
\midrule

\multirow{12}{*}{\textit{Dictionary}} &
Avoid using a pattern on your keyboard. \\
\cmidrule{2-2}
 & Have more variety in the characters you choose. \\
\cmidrule{2-2}
 & Avoid repeating sections. \\
\cmidrule{2-2}
 & Don't repeat the same character many times in a row. \\
\cmidrule{2-2}
 & Avoid strings of characters commonly found in passwords. \\
\cmidrule{2-2}
 & Avoid using very common passwords as part of your own password. \\
\cmidrule{2-2}
 & Don't use common phrases. \\
\cmidrule{2-2}
 & Don't use dictionary words. \\
\cmidrule{2-2}
 & Don't use simple transformations of words or phrases. \\
\cmidrule{2-2}
 & Avoid patterns from the alphabet. \\
\bottomrule
\end{tabular}
\end{table}

\section{Broader Applications of Pedagogical Friction}
\label{appendix_broader_applications}
Pedagogical friction can be applied in a range of scenarios. One example is online privacy consent. Prior work on app privacy labels shows that even short, standardized privacy notices can be hard to understand when they rely on unclear technical terms~\cite{keswani2025user}. Other studies suggest that streamlining notice and consent is not enough on its own. When disclosures are presented as short form labels or as choice interfaces, many people do not spend time reading the details and may skip explanatory text, which limits how informed consent can be in practice~\cite{habib2022evaluating,zhang2024exploring}. In this scenario, pedagogical friction could provide brief explanations of a few key points, followed by a quick comprehension step, such as a short quiz or a simple confirmation prompt, before consent is provided.

Similar issues appear in ongoing privacy controls. Android tracking can occur without meaningful user choice, and privacy permissions can persist after a user has forgotten what they granted or how to revoke it~\cite{kollnig2021fait,prange2024not}. Light forms of pedagogical friction, such as a brief explanation of key trade-offs or a short confirmation step before enabling sensitive access, could help users make these choices more deliberately and revisit them when needed.

\begin{table*}[t]
\centering
\caption{Quiz questions and 20 target rules. The \textit{Target Rule} is shown here for analysis and reporting only; participants saw only the questions.}
\label{tab:survey_questions}
\fontsize{7.5}{12}\selectfont
\begin{tabular}{
    >{\centering\arraybackslash}p{0.6cm} |
    >{\raggedright\arraybackslash}p{7.1cm} 
    >{\raggedright\arraybackslash}p{6.9cm}
}
\toprule
\textbf{Item} & \textbf{Target Rule (tip)} & \textbf{Question: Choose the stronger password} \\
\midrule
1 & Add symbols in unpredictable locations & (1) miSecrt! \quad (2) mi!Secrt \quad (3) equally strong \\
\midrule
2 & Capitalize a letter in the middle & (1) securp1!as \quad (2) securp1!As \quad (3) equally strong \\
\midrule
3 & Mix up your capitalization & (1) 7\#y!5a\&5 \quad (2) 7\#y!5A\&5 \quad (3) equally strong \\
\midrule
4 & Consider inserting digits into the middle & (1) miSecrt!1 \quad (2) miS1ecrt! \quad (3) equally strong \\
\midrule
5 & Move symbols and digits elsewhere in your password & (1) miSecrt! \quad (2) miS!ecrt \quad (3) equally strong \\
\midrule
6 & Avoid using a pattern on your keyboard & (1) qwertyuiop \quad (2) qeryipwuod \quad (3) equally strong \\
\midrule
7 & Have more variety in the characters you choose & (1) abcacabcabc65 \quad (2) abdaczbcavc65 \quad (3) equally strong \\
\midrule
8 & Avoid repeating sections & (1) paspaspas \quad (2) pappaspbs \quad (3) equally strong \\
\midrule
9 & Don't repeat the same character many times in a row & (1) jabsuks!11111 \quad (2) jabsuks!12345 \quad (3) equally strong \\
\midrule
10 & Avoid strings of characters commonly found in passwords & (1) P@ssword!sdrp \quad (2) P@ssward!sdrp \quad (3) equally strong \\
\midrule
11 & Avoid using very common passwords as part of your own password & (1) password122I* \quad (2) possward122I* \quad (3) equally strong \\
\midrule
12 & Don't use site-specific terms in your password & (1) google1!1ehi \quad (2) goo1gle!1ehi \quad (3) equally strong \\
\midrule
13 & Don't use names & (1) pas1!Jake \quad (2) pas1!Jmae \quad (3) equally strong \\
\midrule
14 & Don't use pet names & (1) pas1!Milov \quad (2) pas1!Milpt \quad (3) equally strong \\
\midrule
15 & Don't use common phrases & (1) Secure1!Passwrd \quad (2) Se1cur@Pass!wrd \quad (3) equally strong \\
\midrule
16 & Don't use dictionary words & (1) Apl1!Pas \quad (2) Apple1!Pas \quad (3) equally strong \\
\midrule
17 & Don't use simple transformations of words or phrases & (1) P@ss2Wurd \quad (2) P@st2Wurm \quad (3) equally strong \\
\midrule
18 & Avoid using dates & (1) Oct31st!Yo \quad (2) Octwest!Yo \quad (3) equally strong \\
\midrule
19 & Avoid numerical patterns & (1) 111\$222\$333 \quad (2) 132\$284\$315 \quad (3) equally strong \\
\midrule
20 & Avoid patterns from the alphabet & (1) defg1\$abc \quad (2) dsp1\$aij \quad (3) equally strong \\
\bottomrule
\end{tabular}
\end{table*}

\end{document}